\documentstyle[prl,aps,preprint]{revtex}

\begin{document}
\tightenlines
\draft

\title{Dual symmetry  and the vacuum energy}
\author{ V.I. Tkach$^a$\thanks{E-mail: vladimir@ifug1.ugto.mx},
 J. Socorro $^a$\thanks{E-mail: socorro@ifug4.ugto.mx}
J.J. Rosales $^{a,b}$\thanks{E-mail: juan@ifug3.ugto.mx}, 
and J.A. Nieto $^c$\thanks{E-mail: nieto@uas.uasnet.mx}\\ 
$^a$Instituto de F\'{\i}sica de la Universidad de Guanajuato,\\
Apartado Postal E-143, C.P. 37150, Le\'on, Guanajuato, M\'exico\\
$^b$ Ingenier\'{\i}a en computaci\'on, Universidad del Baj\'{\i}o\\
Av. Universidad s/n Col. Lomas del Sol, Le\'on, Gto., M\'exico\\
$^c$ Facultad de Ciencias F\'{\i}sico-Matem\'{a}ticas, \\
 Universidad Aut\'{o}noma de Sinaloa, C.P. 80010, Culiac\'{a}n, 
Sinaloa, M\'{e}xico. }

\date{\today}

\maketitle
\widetext

\begin{abstract}
In this work we present a new hidden symmetry in gravity for the scale 
factor in the FRW model, for $\rm k=0$. This exact symmetry vanishes
 the cosmological constant. We interpret this hidden symmetry 
as a dual symmetry in the sense that appears in the string theory.
\end{abstract}
\pacs{Pacs No.: 04.20.Jb, 04.90.+e, 98.80.Hw \hspace{7cm}}

\narrowtext

Astronomical observations of the universe indicate that the cosmological 
constant, if it is nonzero,  is very small. The vacuum energy density $\rho_v$ 
multiplied by $\rm 8 \pi G_N=\kappa^2$ (where $\rm G_N$ is the 
Newton constant) is usually called the cosmological constant 
$\rm \Lambda$\cite{W}.
The cosmological data implies that in the present-day vacuum energy density is 
not much greater than the critical density $\rm \rho_c \sim 10^{-48} Gev^4$.

Vacuum energy in quantum field theory is not zero, but it has the value of 
$\rm m^4$, where ``m'' is a characteristic particle physical mass parameter
\cite{CW}. For instance, the masses of intermediate bosons of weak 
interactions is $\rm m_z \sim 10^2 Gev$, which gives 
$\rm \rho \sim 10^8 Gev^4$ and  the observable vacuum energy density 
$\rho_v$ is fantastically small $ \rho_v < 10^{-56} \rho$. This fact 
 could be due to accidental compensation of different contributions to 
$\rho_v$, but there is a little chance for compensation with an accuracy 
of one part in $10^{60}$. It may be  that  the compensation of 
different contributions to $\rho_v$ is secured of a symmetry principle.

The natural candidate is supersymmetry. Experiment shows, however, that this 
symmetry is broken in the observed universe, since the boson and fermion 
masses are different. Therefore, va\-cuun energy is not exactly cancelled. 
In the best case  the contributions to vacuum energy is proportional to 
$\rm m^4_{3/2}$, where $\rm m_{3/2} \sim 10^2-10^3 Gev$ (gravitino 
mass parameter)  describes the scale of supersymmetry breaking\cite{KZP}.

Ideally, we  would like to explain the vanishing of the cosmological 
constant in the observed universe in terms of  an exact 
symmetry principle. In 
the case of early stages of the universe, thus, the symmetry is broken,
 and the 
cosmological constant is not vanishing.
The symmetry must  include a non-trivial new transformations  on the metric
$\rm g_{\mu \nu}(x^\lambda)$, since general coordinate transformations on
the metric $\rm g_{\mu \nu}$ not must constraint the cosmological term
$\rm \sqrt{-g} \Lambda$.

In this work we show, that the minisuperspace formulation allows to have
such a symmetry in any theory of gravitation including  Einstein theory.
We consider a simple model of the universe described by a homogeneous 
and isotropic FRW metric
\begin{equation}
\rm ds^2= - N^2(t) dt^2 + R^2(t) \, d^3 \Omega \, ,
\end{equation}
where $d^3 \Omega$ is the interval on the spatial sector with constant
curvature $\rm k=0, \mp 1$, corresponding to plane, hyperspherical or  
spherical three-space respectively.

The metric is described by a single scale factor $\rm R(t)$ and as the matter
source we shall consider a homogeneous scalar field $\rm \varphi(t)$  which
induce the potential $\rm V(\varphi)$.

The action for FRW interacting with the scalar field $\varphi(t)$ is
described by 
\begin{equation} 
\rm S = \int \left[-\frac{3}{\kappa^2}\frac{R \, {\dot R}^2}{N} 
  + \frac{R^3}{2N} {\dot \varphi}^2 + \frac{3}{\kappa^2}k NR
- NR^3 V(\varphi) \right ]\, dt \, , 
\label{action}
\end{equation}
where $\rm \dot R = \frac{dR}{dt}$, $\rm \dot \varphi = \frac{d\varphi}{dt}$ 
and $\rm \kappa=(8 \pi G_N)^{1/2}$ have length dimensions $\ell$. The 
dimension of the scalar field $\rm \varphi(t)$ is of the form  
 $\ell^{-1}$, while the potential  $\rm V(\varphi)$ has dimension 
$\ell^{-4}$.  We assume units in which $\rm c=\hbar=1$.
 
It turns out that the action S is   invariant under the time
reparametrization $\rm t \to t^\prime = t + a(t)$, if the variables
$\rm N(t), R(t)$ and $\rm \varphi(t)$ are transformed as
\begin{equation}
\delta N = (a N \dot ), \qquad \delta R = a \dot R, \qquad \delta \varphi=
a \dot \varphi \, .
\label{transformation1}
\end{equation}
In fact, under the transformation (\ref{transformation1}) the
action (\ref{action}) becomes 
\begin{equation}
\rm \delta S = \int (a L \dot ) \, dt ,
\end{equation}
where L is the corresponding Lagrangian. So, up to a total derivative the
action S is invariant under the transformation (\ref{transformation1}).

We can see, that the first and the second terms in the action (\ref{action}),
which are the kinetic terms for the scale factor $\rm R(t)$ and the scalar 
field $\rm \varphi(t)$ respectively are invariant under the following 
transformations, that in what follow we call dual transformation
\begin{eqnarray}
{\rm R(t)} &\rightarrow & {\rm R^\prime (t)= \frac{\kappa^2}{ R^(t)}}, \qquad 
{\rm N(t)}  \rightarrow  {\rm N^\prime (t) = \frac{\kappa^6 N(t)}{ R^6(t)}}, 
\nonumber\\
{\rm \varphi(t)}&  \rightarrow & {\rm \varphi(t)= \varphi(t)} . 
\label{transformation2}
\end{eqnarray}
It is worth mentioning it that these transformations can be 
understood as the
analogue of the T-duality in string theory\cite{GiPoRa}. Since the duality of
dualities \cite{Du} assumes that S and T dualities can be interchanged, it 
appears interesting to see what could be the S-dual transformation
of (\ref{transformation2}). In this direction our work may be related to
Witten's paper \cite{Wi}.  Explicitely,
the target space in the modular transformations of the 
string contains the well-known dual transformations $\rm r \to \frac{1}{r}$,
where ``r'' is the ``radius'' of the internal six-dimensional space of 
the string \cite{muchos}.

The term $\rm \frac{3}{\kappa^2} kNR$ is invariant under the transformation
(\ref{transformation2}) only for the plane three-space, $\rm k=0$.
Finally, the last term in the action (\ref{action}), $\rm -NR^3 V(\varphi)$,
is the effective cosmological constant term and defines the contributions to 
cosmological constant from the potential $\rm V(\varphi)$. If $\rm V(\varphi)$
vanishes at $\rm \varphi = < \varphi >= \varphi_0$, for example for a  
potential of the form $\rm V= \lambda (\varphi^2 - a^2 )^2$, 
the action (\ref{action}) is invariant under the dual transformation 
(\ref{transformation2}), while for the vacuum energy $\rm V(\varphi=0) = a^4 
\not= 0$ the dual symmetry is broken and the cosmological constant is
non-vanishing.

Let us find the canonical Hamiltonian of the model. The momenta 
conjugate to R and $\varphi$ are defined in the usual way
 
\begin{equation}
\rm \Pi_R = - \frac{6}{\kappa^2} \frac{R \dot R}{N}, \qquad
\Pi_\varphi = \frac{R^3 \dot \varphi}{N},
\label{momentum}
\end{equation}
and under the dual transformations (\ref{transformation2})
we have the following relations between the old and the transformed momenta
\begin{equation}
\rm \Pi_R \to \Pi^\prime_R = - \frac{1}{\kappa^2} R^2 \Pi_R, \qquad
\Pi_\varphi \to \Pi_\varphi^\prime = \Pi_\varphi \, .
\label{transformation3}
\end{equation}

With these relations we find, that the canonical Poisson brackets are
invariant under dual transformations (\ref{transformation2}):
\begin{equation}
\rm \left\{ R, \Pi_R \right \} = \left\{ R^\prime, \Pi_R^\prime \right \}=1,
\qquad \left\{ \varphi, \Pi_\varphi \right \} = 
\left\{ \varphi^\prime, \Pi_\varphi^\prime \right \} = 1 \, .
\label{poisson}
\end{equation}

Thus,  the Hamiltonian can be calculated in the usual way. We have the 
classical canonical Hamiltonian
\begin{equation}
\rm H_c = N H = N \left[ 
- \frac{\kappa^2}{12 R} \Pi^2_R +\frac{1}{2R^3}\Pi^2_\varphi + R^3 V(\varphi)
\right] ,
\label{hamiltonian}
\end{equation}
where H is the Hamiltonian of the system. This form of the canonical 
Hamiltonian explains the fact, that the lapse function N is a Lagrange
multiplier, which enforces the only first-class constraint $\rm H=0$. 
This constraint, of course expresses the invariance of the action 
(\ref{action}) under reparametrization transformations.

We note that for the case $\rm k=0$ and for the effective cosmological
 term  $\rm R^3 V(\varphi_0) = 0$ the canonical Hamiltonian  is invariant 
under dual transformations (\ref{transformation2}). 
In fact, since under (\ref{transformation1}) the classical constraint H 
transform as $\rm H \to H^\prime = H R^6/\kappa^6$, 
we get $\rm N^\prime H^\prime = NH$.

According to the Dirac's constraints Hamiltonian quantization procedure, 
the wave function is annihilated
by operator version of the classical constraint. In the usual fashion, the 
canonical momemta are replaced by operators
$\rm \hat \Pi_R= - i \frac{\partial}{\partial R}, 
\hat \Pi_\varphi = - i \frac{\partial}{\partial \varphi}$.

It turns out that the commutators of the quantum operators 
$\rm \left[R, \hat \Pi_R \right ] = i$, $\rm \left[\varphi, \hat \Pi_\varphi 
\right] = i$ are also  invariants under the dual transformations.

In order to find a correct quantum expression for the Hamiltonian we must 
always  considerate factor ordering ambiguities. This is truth in our case
because the operator Hamiltonian contains the product
non-commutating operator R and $\rm \hat \Pi_R$. Thus, of the first term in 
the classical Hamiltonian H we consider the following operator form \cite{M}
\begin{equation}
\rm \frac{\kappa^2}{12} R^{-p-1} \frac{\partial}{\partial R} R^p 
\frac{\partial}{\partial R} = \frac{\kappa^2}{12} \left (
\frac{1}{R} \frac{\partial^2}{\partial R^2} + \frac{p}{R^2} 
\frac{\partial}{\partial R}  \right) ,
\label{ordering}
\end{equation} 
where p is a real parameter, that measures the ambiguity in the factor 
ordering \cite{HH} in the first term of (\ref{hamiltonian}).

The quantum Hamiltonian has the form
\begin{eqnarray}
{\rm \hat H }&=&{\rm \frac{\kappa^2}{12} R^{-p-1} \frac{\partial}{\partial R} 
R^p \frac{\partial}{\partial R} - 
\frac{1}{2R^3} \frac{\partial^2}{\partial \varphi^2} + R^3 V(\varphi)},
\nonumber\\
&=& {\rm \frac{\kappa^2}{12 R} \frac{\partial^2}{\partial R^2}  
+ \frac{\kappa^2\, p}{12 R^2} \frac{\partial}{\partial R} - 
\frac{1}{2R^3} \frac{\partial^2}{\partial \varphi^2} + R^3 V(\varphi)}.
\label{new}
\end{eqnarray}
On the other hand, under duality transformations the Hamiltonian H becomes 
\begin{equation}
\rm \hat H^\prime= \frac{R^6}{\kappa^6}  \hat H,
\end{equation}
Thus, we may assume the final form
\begin{equation}
\rm \hat H =\frac{\kappa^2}{12} R^{p-3} \frac{\partial}{\partial R} R^{-p+2} 
\frac{\partial}{\partial R} - 
\frac{1}{2R^3} \frac{\partial^2}{\partial \varphi^2} +\frac{\kappa^{12}}{R^9}
 V(\varphi),
\end{equation}
It is straightforward to show, that $\rm \hat H$ is dual invariant only if
the parameter $\rm p=1$ and $\rm V(\varphi_0)=0$.
If $\rm V(\varphi_0) \not= 0$ the last term in (\ref{new}) is broken
under the dual symmetry.

The physical states $\rm |\Psi>$ are those that are annihilated 
by $\rm \hat H$:

\begin{equation}
\rm \hat H |\Psi> =  \left[\tilde R^2 \frac{\partial^2}{\partial \tilde R^2}
+ p  \tilde R\frac{\partial}{\partial \tilde R} 
-6 \frac{\partial^2}{\partial \tilde \varphi^2}
- 36 k \tilde R^4 + 12 \lambda \,  \tilde R^6 (\tilde \varphi^2- \tilde a^2)^2
 \right ] |\Psi> =0 .
\label{quantum-hamiltonian}
\end{equation}
This equation can be identified with the Wheeler-DeWitt equation for a 
minisuperspace models. Clearly, (\ref{quantum-hamiltonian}) has
many different solutions, and one of the most fundamental questions facing
us is which of these solutions actually describes our universe.

Let us consider the following ansatz 

\begin{eqnarray}
a& ) & \,\, {\rm \Psi(\tilde R,\varphi_0) = 
\Psi(\frac{1}{\tilde R}, \tilde \varphi_0), \qquad 
V(\tilde \varphi_0= \tilde a)=0,}\nonumber\\
b& ) &\,\, V(\tilde \varphi=0)=\tilde a^4, {\rm dual~breaking~ invariance}, 
\qquad
\Psi(\tilde R) \not= \Psi(\frac{1}{\tilde R}, \tilde \varphi=0). 
\end{eqnarray}
Here, we consider the scale factor and the scalar field, 
dimensionless; 
$\rm \tilde R= \frac{R}{\kappa}$, $\tilde \varphi= \kappa \varphi$, 
$\rm \tilde a = \kappa a$.
Considering these two cases,
we shall find the exact solution to the quantum equation 
(\ref{quantum-hamiltonian}), for any p. 

 The case a) means that we need to ensure the invariance of the
 Hamiltonian (\ref{new})  under dual transformation  

\begin{eqnarray}
{\rm \tilde R(t)} &\rightarrow & {\rm \frac{1}{ \tilde R^(t)}}, \qquad 
{\rm N(t)}  \rightarrow  {\rm \frac{ N(t)}{\tilde R^6(t)}}, \nonumber\\
{\rm \tilde \varphi(t)}&  \rightarrow & {\rm \tilde \varphi(t)} 
\qquad {\rm k}:=0, \quad
{\rm and} \quad{\rm  p:=1.} 
\label{transformationn}
\end{eqnarray}
Note, that in this case
the values of the parameter $\rm k$, the 
factor ordering p, and  the potential for the scalar field are fixed.
Thus, under the duality transformations (\ref{transformationn}) we have

\begin{equation}
\rm \hat H |\Psi> =  \left[\tilde R^2 \frac{\partial^2}{\partial \tilde R^2}
+\tilde   R\frac{\partial}{\partial \tilde R} 
-6 \frac{\partial^2}{\partial \tilde \varphi^2} \right ] |\Psi> =0, 
\label{quantum-ha}
\end{equation}

\begin{equation}
\Psi(\tilde R,\tilde \varphi)= \left \{  
 \begin{array}{l}
{\rm A\, \left(\tilde   R^\nu 
+  \tilde R^{-\nu} \right) e^{\pm \frac{ \nu}{\sqrt{6}}\tilde \varphi} }; 
\qquad {\rm m=-\nu^2<0,}\cr 
{\rm B cos \left(\nu Ln \tilde R \right) 
\,  e^{\pm i\frac{\nu}{\sqrt{6}} \tilde \varphi} }; 
\qquad {\rm m=\nu^2} \,  > \, 0 \, ,
\end{array} 
\right.
\label{solution-a} 
\end{equation}
where $\rm A, B$ and $\nu$ are constants of integration. 

Now, for the case b) the factor ordering is not  fixed and we obtain
the solution
\begin{equation}
\rm  \Psi(\tilde R,\tilde \varphi) = \tilde R^{\frac{1-p}{2}} Z_\nu\left(
\frac{\sqrt{2}\tilde a^2}{3} \tilde R^3 \right)\,  e^{\pm im \tilde \varphi}, 
\qquad 
\nu=\frac{1}{6} \sqrt{(1-p)^2- (2m)^2}\, , 
\label{exact-solution-general}
\end{equation}
where $\rm Z_\nu\left(\frac{\sqrt{2}\tilde a^2}{3} \tilde R^3 \right)$ is the 
Bessel function defined as $\rm A J_\nu + B Y_\nu$, where $\nu$ is a real 
(or imaginary \cite{Dun}) number with a real  argument 
$\frac{\sqrt{2}\tilde a^2}{3} \tilde R^3 $.

It is straightforward to generalize our procedure to
all Bianchi type 
cosmological models. In the proposed framework it is also possible to include
the supersymmetric minisuperspace models\cite{TkRO}. 
Thus, as further research it may be interesting to consider the close 
minisuperspace models.

Furthermore, these simple dual transformations may be applied to the metric 
$\rm g_{\mu \nu}(x^\lambda)$  in the ADM formalism \cite{ADM}. 
According to the ADM prescription, of general relativity, they are 
considered a 
slicing of the space-time by family of space-like
hypersurfaces labeled by a parameter t. The space-time metric 
$\rm g_{\mu \nu}(x^\lambda)$ is decomposed into lapse $\rm N(t,x^k)$, 
shift $\rm N_i(t, x^k)$ and 
the three-metric of the slice $\rm h_{ij}(t,x^k)$. Thus, these quantities
under dual transformations have the following form

\begin{equation}
\rm N \to N^\prime = \frac{N}{h}, \qquad N_i \to N_i^\prime = 
\frac{N_i}{h^{2/3}}, \qquad h_{ij} \to h_{ij}^\prime = 
\frac{h_{ij}}{h^{2/3}},
\label{transformation4}
\end{equation}
where $\rm h= det(h_{ij})$.

A gravitational vacuum polarization correction to effective action\cite{DDI} 
must induce new terms which will be not invariant under dual transformations  
(\ref{transformation4}).  In fact, even if
$\rm \sqrt{-g} V(\varphi)$  vanishes at 
$< \varphi >=\varphi_0$ quantum correction to $\rm V(\varphi)$ would generate
non-zero contribution $\rm \sqrt{-g} V_{eff}(\varphi_0)$ and the effective 
action is not 
invariant under dual transformations (\ref{transformation4}).
Therefore, the condition for exact dual symmetry leads to the cancelation of 
these two contributions of the vacuum energy.

\acknowledgments
We are grateful to A. Cabo, J. Lucio M, D. Kazakov,I. Lyanzuridi,
L. Marsheva and O. Obreg\'on for their interest in this paper.
This work was partially 
supported by  CONACyT, grant No. 3898P--E9608.

\end{document}